\documentclass[conference,10pt]{IEEEtran}
\usepackage[T1]{fontenc}
\usepackage[utf8]{inputenc}
\usepackage{mathptmx}
\usepackage{amsmath,amssymb}
\usepackage{graphicx}
\usepackage{booktabs}
\usepackage{siunitx}
\usepackage[hidelinks]{hyperref}
\usepackage{microtype}
\usepackage{caption}
\usepackage{stfloats}
\usepackage{tikz}
\usetikzlibrary{arrows.meta,positioning,calc,fit,backgrounds,decorations.pathmorphing,patterns}

\graphicspath{{figs/}}
\hypersetup{
  pdftitle={A Multimode-Fiber Speckle Sensing Node with a Press-Robust Quantized Network for Liquid-Cooling Thermal Monitoring},
  pdfauthor={Xuezhi Lou, Weipeng Zhang, Zhibai Zhang, Liyuan Ren},
  pdfkeywords={fiber speckle, multimode fiber, liquid cooling, quantized neural network, FPGA, edge inference}
}

\begin{document}

\newcommand{\blfootnote}[1]{%
  \begingroup
  \renewcommand{\thefootnote}{}%
  \footnotetext{#1}%
  \endgroup}

\title{A Multimode-Fiber Speckle Sensing Node with a\\Press-Robust Quantized Network for\\Liquid-Cooling Thermal Monitoring}

\author{\IEEEauthorblockN{Xuezhi Lou\textsuperscript{1,*}, Weipeng Zhang\textsuperscript{1,*}, Zhibai Zhang\textsuperscript{1,*}, and Liyuan Ren\textsuperscript{2,\dag}}\\
\IEEEauthorblockA{\textsuperscript{1}Glasgow College, University of Electronic Science and Technology of China, Chengdu, China\\
\textsuperscript{2}Institute of Applied Physics and Computational Mathematics, Beijing, China\\
\textsuperscript{*}Equal contribution. \textsuperscript{\dag}Corresponding author.}}

\maketitle
\vspace*{-11pt}
\blfootnote{\textsuperscript{\dag}Corresponding author, ren\_liyuan@iapcm.ac.cn}

\begin{abstract}
Liquid-cooled AI and HPC racks concentrate heat at cold plates and manifolds, below the granularity of facility sensors. We present a self-contained edge node that turns one multimode fiber into a distributed thermal-event sensor over nine pipe segments, with a 450~nm laser, a camera and a Zynq UltraScale+ device that captures, infers, displays and reports over MQTT. The speckle decorrelates within tens of seconds at constant temperature, and a press can change it as much as a 1~$^\circ$C rise. The node averages each 20~s window and feeds two planes, the mean pattern and a 19-dimensional temporal-feature map, to a 61.6k-parameter int8 convolutional neural network (CNN) that treats presses as hard negatives and runs in the programmable logic at \SI{565}{\micro\second} per frame. On a 59-session, 70\,148-window dataset, session-held-out validation gives a pooled AUC of 0.961 against 0.876 for decorrelation, no alarm during the hold or the 45~s rebound in 23 press trials against 6 and 16, and a mean per-fold step detection of 83.8\% at 0.4 false alarms per hour.
\end{abstract}


\section{Introduction}
\IEEEPARstart{P}{ower} density has pushed data-center cooling to chip-level liquid cooling \cite{kong2024}. Published energy data for Frontier exposes only loop-level temperatures \cite{sun2024}, and cooling benchmarks use the same aggregates \cite{naug2025}. The sensors that reveal a blocked cold plate sit at the manifold and plate level, where every point costs wiring or a coolant penetration.

A multimode fiber is a good transducer, since a temperature change anywhere alters the modal phases and the exit speckle \cite{redding2013}. Speckle has been used for strain \cite{murray2019}, curvature \cite{fujiwara2022}, wavelength \cite{metzger2017,bruce2019} and, with learned demodulation, temperature \cite{yang2025,velez2026,gao2025}; distributed sensing \cite{bao2012,gemeinhardt2024} resolves meters, not cold plates. A node needs one fiber, one camera and a board.

Two properties defeat naive demodulation. The speckle drifts, the correlation falling to 0.6--0.7 after 60~s at constant temperature \cite{metzger2017,bruce2019}, and a press can decorrelate it as much as a 1~$^\circ$C rise. We contribute a complete node, a 59-session dataset with press and acoustic distractors, a two-plane Speckle Pattern Inference Engine (SPINE) at 2.6-million-multiply-accumulate (MMAC) whose inference runs in the programmable logic (PL), and comparisons against feature-based, backbone and foundation models.

\begin{figure}[!t]
\centering
\includegraphics[width=\columnwidth]{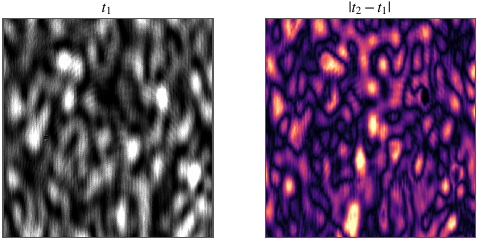}
\caption{A speckle frame (left) and its residual against a frame 30~s later (right), in pseudo-color; $\Delta T=0.57~^\circ$C and $1-r=0.51$.}
\label{fig:speckle}
\end{figure}

\section{Device and Architecture}
A 450~nm laser feeds 3~m of multimode fiber wound on nine heated pipe segments. Sixteen 12~V/5~W heaters are switched by metal-oxide-semiconductor field-effect transistors (MOSFETs), and thirteen DS18B20 probes are recorded on an eighteen-channel frame, nine as feedback. An STM32F103 runs the 1~Hz proportional-integral-derivative (PID) loop and de-energizes all heaters on any fault. An OV9281 camera images the exit facet at $640\times800$ RAW8 over one MIPI lane; exposure is locked and five frames within 0.5~s are median-fused to 1~Hz, which keeps brightness from becoming a shortcut.

\begin{figure*}[!t]
\centering
\input{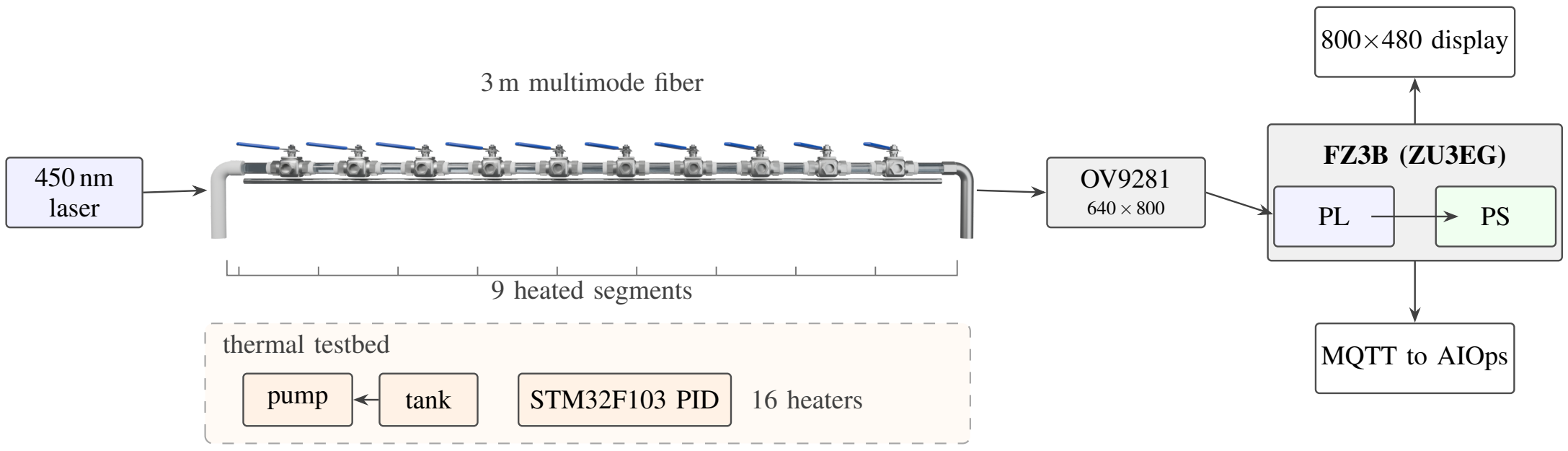}
\caption{The sensing node. The thermal testbed is used for collection and calibration only.}
\label{fig:device}
\end{figure*}

Fig.~\ref{fig:device} shows the node; the PL receives the Mobile Industry Processor Interface (MIPI) D-PHY stream, parses Camera Serial Interface 2 (CSI-2) with error-correcting-code (ECC) and cyclic-redundancy-check (CRC) checks, and writes frames to DDR4 by video direct memory access (VDMA) at 149.8~fps. The SPINE core reads $2\times56\times56$ over Advanced eXtensible Interface (AXI) direct memory access (DMA), keeps every feature map on chip and pipelines frames between layers. The processing system (PS) runs the window buffer, the nineteen temporal features, the alarm rule and MQTT; a PL compositor builds the $800\times480$ screen.

\begin{figure}[!t]
\centering
\includegraphics[width=\columnwidth]{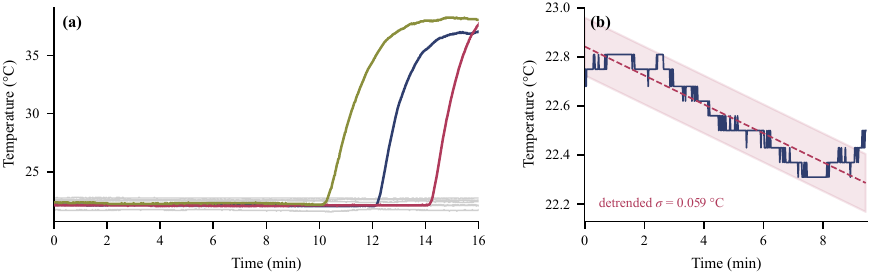}
\caption{Thermal testbed. (a)~Stepped heating of three segments. (b)~Quiescent channel, detrended $\sigma=0.059~^\circ$C.}
\label{fig:bench}
\end{figure}

The testbed reproduces stepped heating, and a quiescent channel shows a detrended noise of 0.059~$^\circ$C (Fig.~\ref{fig:bench}). At 100~MHz a frame takes \SI{565}{\micro\second} in simulation, \SI{1.12}{ms} for the first and \SI{1.33}{ms} on a host round trip. The design uses 28.8\% look-up tables (LUTs), 14.2\% flip-flops (FFs), 24.8\% block random-access memory (BRAM) and 30.0\% digital-signal-processing (DSP) slices at $+2.55$~ns worst slack; on-chip power is \SI{2.99}{W}, the inference core about \SI{0.21}{W}, or \SI{0.29}{mJ} per \SI{1.33}{ms} inference.

\section{Dataset}
A session is one recording under one configuration and is the unit of splitting. A window is heating when any feedback channel moves by at least 1.0~$^\circ$C within 120~s, in either direction, and normal when no channel moves more than 0.25~$^\circ$C and no setpoint changed within 300~s. Press trials are cued, and the interval runs to 45~s after release to cover the acrylic rebound.

The corpus has 59 sessions and 57\,413 labelled windows in five laser power-on domains.

Let $Z_t$ be the normalized frame at time $t$ and $A,B$ two window means. Their similarity is
\begin{equation}
r(A,B)=\frac{\sum_i (A_i-\bar A)(B_i-\bar B)}
{\sqrt{\sum_i (A_i-\bar A)^2}\sqrt{\sum_i (B_i-\bar B)^2}},
\end{equation}
and the decorrelation is $d=1-r$. The detection task is to decide, from windows alone, whether any of the nine feedback points moved by at least $1~^\circ$C within 120~s.

\begin{figure*}[!t]
\centering
\includegraphics[width=0.84\textwidth]{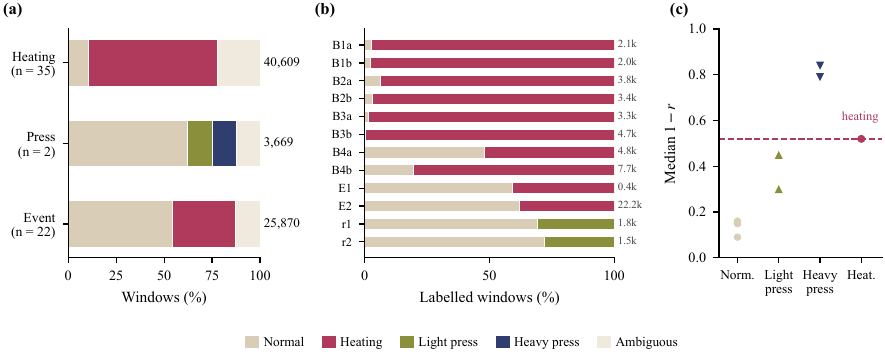}
\caption{Dataset characteristics. (a)~Window composition by session type. (b)~Labelled windows per fold. (c)~Median window $1-r$ by class.}
\label{fig:data}
\end{figure*}

\begin{figure*}[!t]
\centering
\includegraphics[width=0.82\textwidth]{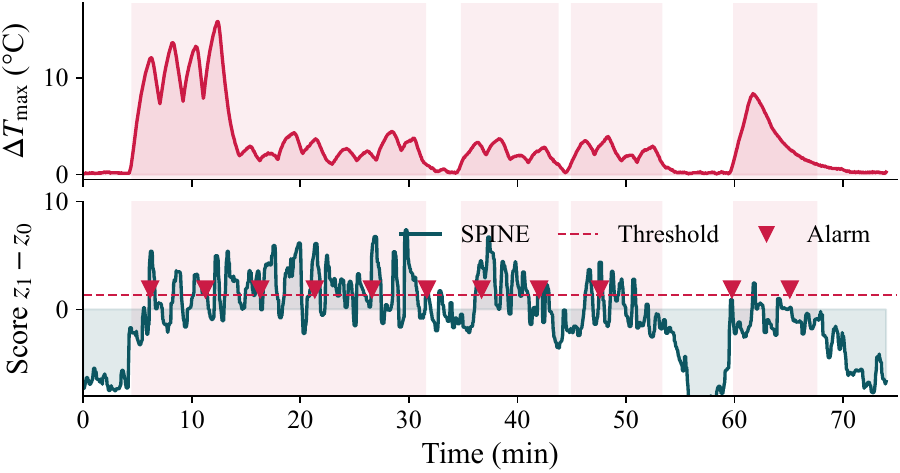}
\caption{Alarm process of a 74-min heating session. Top, maximum temperature change, with shaded bands where the rise exceeds 1~$^\circ$C. Bottom, the score $z_1-z_0$ with 11-point median smoothing; the dashed line is the 99th-percentile threshold and triangles mark raised alarms.}
\label{fig:ts}
\end{figure*}

Two properties of the data set the task. Time drift: at constant temperature the frame correlation falls from 0.99 at 1~s to 0.45 at 60~s and 0.23 at 300~s, while a heating session decays faster. Amplitude overlap: window $1-r$ separates normal from heating, but light presses sit between the two and heavy presses reach the heating level (Fig.~\ref{fig:data}c), so a heavy press is as decorrelating as a 1~$^\circ$C rise and $1-r$ alone gives an area under the receiver operating characteristic curve (AUC) of 0.53. Together they motivate comparing window means and learning the temporal signature rather than thresholding amplitude. Fig.~\ref{fig:ts} shows a 74-min heating session: the score follows every step and the detector raises one alarm per step.

\section{Method}
Frames are row-binned to $400\times640$, divided by a Gaussian-blurred copy ($\sigma=40$~px) and standardized. Plane~0 is the window mean
\begin{equation}
\bar Z_w=\frac{1}{|w|}\sum_{t\in w} Z_t,
\end{equation}
cropped to $392\times392$, block-averaged to $56\times56$ and quantized with scale 40 and zero point 128. Plane~1 holds nineteen temporal features, six multiscale decorrelations, three adjacent-frame statistics, a 10~s mean ratio, four within-window deviations, recovery, monotonicity and brightness and contrast changes, quantile-normalized and mapped to 8-bit by
\begin{equation}
q(x)=\mathrm{clip}\!\left(\mathrm{round}\!\left(\frac{x-\ell}{h-\ell}\,255\right),0,255\right),
\end{equation}
with $\ell,h$ the 2nd and 98th training percentiles, then tiled to $56\times56$.

The network is four $3\times3$/2 convolutions with 16/32/64/64 channels, a $4\times4$ global average pooling (GAP) layer and two fully connected (FC) layers, giving 61.6k parameters and 2.6-million-multiply-accumulate (MMAC). Channel~0 is normal, 1 heating, 4--8 a five-dimensional embedding. Training uses class-weighted cross entropy; presses are duplicated six times as hard negatives, the embedding is regularized by Deep~SAD \cite{ruff2020}, and the last 8 of 38 epochs are quantization-aware \cite{jacob2018} with int8 weights and 12-bit activations. The score is
\begin{equation}
s_t=z_1(t)-z_0(t),
\end{equation}
thresholded at the 99th percentile $\theta$ of training-fold normal windows; an alarm is raised when
\begin{equation}
a_t=\mathbb{1}\left[s_t>\theta,\; s_{t-1}>\theta,\; s_{t-2}>\theta\right],
\end{equation}
evaluated every 2~s with a 300~s cooldown.

\section{Experiments}
All results use twelve-fold session-held-out validation; thresholds are set on training folds. Baselines are $1-r$, gradient boosting and logistic regression on the same features.

\begin{figure*}[!t]
\centering
\includegraphics[width=0.86\textwidth]{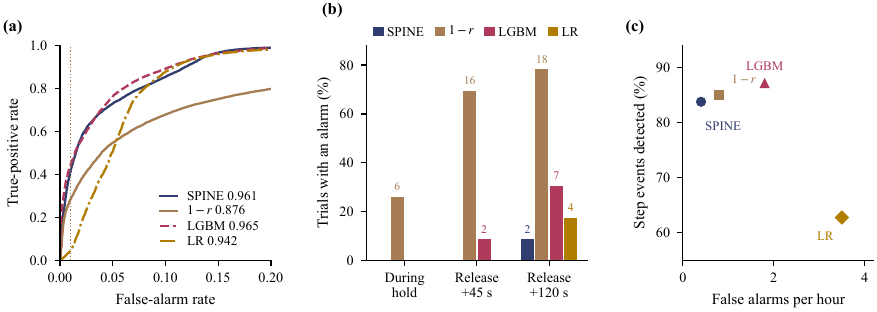}
\caption{(a)~Pooled ROC over the first 20\% of the false-alarm axis. (b)~Press-trial alarms. (c)~Event-level trade-off.}
\label{fig:res}
\end{figure*}

\begin{table*}[t]
\centering
\caption{Comparison under the same twelve-fold, session-held-out split. AUC is pooled; detection is the mean per-fold step-event rate and delay its median, reported for event-level runs.}
\label{tab:all}
\footnotesize
\renewcommand{\arraystretch}{0.95}
\setlength{\tabcolsep}{7pt}
\begin{tabular}{@{}lcccc@{}}
\toprule
Method & AUC & Detected (\%) & Delay (s) & Platform \\
\midrule
SPINE (ours) & 0.961 & 83.8 & 179 & PL \\
MLP (66.7k)  & 0.957 & --   & --  & GPU \\
TinyCNN (61.6k) & 0.777 & -- & -- & GPU \\
$1-r$        & 0.876 & 85.0 & 148 & PS \\
LGBM         & 0.965 & 87.2 & 174 & CPU \\
LR           & 0.942 & 62.8 & 214 & CPU \\
TabICL       & 0.970 & --   & --  & GPU \\
TabPFN    & 0.970 & --   & --  & GPU \\
\bottomrule
\end{tabular}
\end{table*}

Table~\ref{tab:all} and Fig.~\ref{fig:res} summarize. Pooled AUC rises from 0.876 to 0.961, recall at 1\% false alarms from 28.4\% to 41.3\%, and the press fraction above the 90\%-recall threshold falls from 77.8\% to 10.8\%. Gradient boosting reaches 0.965 and 44.6\%, on par with the SPINE but requiring a CPU pipeline; logistic regression reaches 0.942 yet only 4.6\% at 1\% false alarms. The SPINE is the only compared model whose inference runs in the PL. Parameter matching isolates the gain. A 61.6k-parameter CNN on the same planes reaches 0.777, and a 66.7k-parameter MLP on the features reaches 0.957; neither rejects presses.

At the event level the SPINE detects 83.8\% per fold at 0.4 false alarms per hour, against 85.0\% at 0.8 for $1-r$, 87.2\% at 1.8 for boosting and 62.8\% at 3.5 for logistic regression, matching detection at half the false alarms. The median delay is 179~s, with a 2.9~$^\circ$C change at the alarm. Over 23 press trials the deployed detector alarms 6 times during the hold and 16 times within 45~s of release, while the SPINE alarms in none of the 23 during the hold or the rebound window.

Table~\ref{tab:all} compares all methods under the same split. SPINE reaches 0.961 and needs 2.6~MMAC per inference, or 1.33~ms on the board; gradient boosting and logistic regression reach 0.965 and 0.942 with microsecond CPU inference, and TabICL \cite{tabicl2025} and TabPFN \cite{tabpfn2025} reach 0.970 at sub-millisecond latency on a GPU. SPINE is the only model whose inference runs on the board and the only one trained to reject presses.

\section{Discussion}
SPINE gains its margin from two choices. The window mean removes the fast component of laser drift, and the nineteen temporal features expose the rise-versus-transient signature that a single decorrelation value lacks; sixfold press duplication then pushes transients to the non-heating side. A detector at 2.6~MMAC rejects presses and holds 0.4 false alarms per hour, while the foundation models need a host GPU for a 0.009 AUC gain. A control experiment with ResNet-18 and MobileNetV3-S trained on the same $56\times56$ planes reached only 0.57--0.66 AUC at 5--127~MMAC, so raw capacity does not replace the two-plane design.

\section{Conclusion}
We described a multimode-fiber speckle node, a 59-session dataset with press and acoustic distractors, and SPINE. It removes press-induced alarms during the hold and rebound, holds 0.4 false alarms per hour at 83.8\% step detection, and reaches a pooled AUC of 0.961.

\bibliographystyle{IEEEtran}
\bibliography{refs}

\end{document}